\documentclass[aps,pra,twocolumn,amsmath,amssymb,nofootinbib,showpacs,superscriptaddress]{revtex4-2}
\usepackage[english]{babel}
\usepackage[normalem]{ulem}
\usepackage{latexsym}
\usepackage{graphics}
\usepackage{graphicx}
\usepackage{epsfig}
\usepackage{color}
\usepackage{bm}
\usepackage{amsmath}
\usepackage{amssymb}
\usepackage{amsthm}
\usepackage{dcolumn}
\usepackage{bm}
\usepackage{float}
\usepackage{hyperref}
\usepackage{color}
\usepackage{epstopdf}
\usepackage{cleveref}
\usepackage[svgnames]{xcolor}
\usepackage{comment}
\usepackage[symbol]{footmisc}
\usepackage{soul}

\hypersetup{hidelinks,colorlinks=true,allcolors=DarkBlue}

\newcommand{\teal}[1]{{\color{Black} #1}}

\newcommand{\abs}[1]{\left| #1 \right|}
\newcommand{\ket}[1]{\left | #1 \right \rangle}

\theoremstyle{remark}

\begin{document}

\preprint{APS/123-QED}

\title{Reconstructing complex states of a 20-qubit quantum simulator}

\author{Murali K. Kurmapu}
\affiliation{University of Calgary, Calgary AB T2N 1N4, Canada}
\affiliation{Department of Physics, University of Oxford, Oxford OX1 3PG, UK}

\author{V.V. Tiunova}
\affiliation{Russian Quantum Center, Skolkovo, Moscow 143025, Russia}

\author{E.S. Tiunov}
\affiliation{Quantum Research Centre, Technology Innovation Institute, Abu Dhabi, UAE}

\author{Martin Ringbauer}
\affiliation{Institut f\"{u}r Experimentalphysik, Universit\"{a}t Innsbruck, Technikerstrasse 25, 6020 Innsbruck, Austria}

\author{Christine Maier}
\affiliation{Alpine Quantum Technologies GmbH, 6020 Innsbruck, Austria}

\author{Rainer Blatt}
\affiliation{Institut f\"{u}r Experimentalphysik, Universit\"{a}t Innsbruck, Technikerstrasse 25, 6020 Innsbruck, Austria}
\affiliation{Alpine Quantum Technologies GmbH, 6020 Innsbruck, Austria}
\affiliation{Institut f\"{u}r Quantenoptik und Quanteninformation, \"{O}sterreichische Akademie der  Wissenschaften, Otto-Hittmair-Platz 1, 6020 Innsbruck, Austria}

\author{Thomas Monz}
\affiliation{Institut f\"{u}r Experimentalphysik, Universit\"{a}t Innsbruck, Technikerstrasse 25, 6020 Innsbruck, Austria}
\affiliation{Alpine Quantum Technologies GmbH, 6020 Innsbruck, Austria}

\author{Aleksey K. Fedorov}
\affiliation{Russian Quantum Center, Skolkovo, Moscow 143025, Russia}
\affiliation{National University of Science and Technology ``MISIS'', 119049 Moscow, Russia}

\author{A.I. Lvovsky}
\email{Corresponding author: Alex.Lvovsky@physics.ox.ac.uk}

\affiliation{Department of Physics, University of Oxford, Oxford OX1 3PG, UK}

\begin{abstract}
A prerequisite to the successful development of quantum computers and simulators is precise understanding of  physical processes occurring therein, which can be achieved by measuring the quantum states they produce. 
However, the resources required for traditional quantum-state estimation  scale exponentially with the system size, highlighting the need for alternative approaches. Here we demonstrate an efficient method for reconstruction of significantly entangled multi-qubit quantum states. 
Using a variational version of the matrix product state ansatz, we perform the tomography (in the pure-state approximation) of quantum states produced in a 20-qubit trapped-ion Ising-type quantum simulator, using the data acquired in only 27 bases with 1000 measurements in each basis. 
We observe superior state reconstruction quality and faster convergence compared to the methods based on neural network quantum state representations: 
restricted Boltzmann machines and feedforward neural networks with autoregressive architecture. 
Our results pave the way towards efficient experimental characterization of complex states produced by the quench dynamics of many-body quantum systems.
\end{abstract}
\maketitle

Quantum computers and  simulators have the potential to solve computationally difficult problems by utilizing the controllable dynamics of engineered quantum systems~\cite{feynman1982simulating,feynman1986quantum,lloid1996universal, cirac2012goals}. 
The latest programmable quantum simulators, in particular those based on cold atoms~\cite{gross2017quantum} and trapped ions~\cite{Blatt2012}, 
have been used to explore quantum many-body phenomena close to the threshold of what is possible to simulate using classical resources~\cite{mazurenko2017cold,bernien2017probing,keesling2019quantum,qubits20,omran2019generation,ebadi2021quantum,bluvstein2020controlling,scholl2021quantum}. 
The development of such technology hinges on our understanding of the physical processes within the quantum device at the level of both individual qubits and the entire system. 
In order to deepen this understanding, we ought to be able to measure, characterize, and reconstruct the states produced in that device --- that is, to master the art known as quantum-state tomography (QST)~\cite{paris2004quantum}.  

QST becomes increasingly challenging due to the exponential scaling of the number of required measurements and computational resources with the size (number of qubits) of a quantum system.
This gives rise to a qualitatively new requirement: to reconstruct the state using incomplete measurement data in a computationally efficient manner.
Addressing this challenge is of importance for further progress in quantum simulation and, as a consequence, understanding complex and non-trivial quantum many-body phenomena. 

Large quantum systems can be tackled by means of so-called variational approaches. 
Their idea is to represent a high-dimensional quantum state with an ansatz that is dependent on much fewer parameters than the dimension of the Hilbert space under consideration. 
By using standard iterative optimization methods, these parameters can be adjusted such that the corresponding quantum state becomes optimal with respect to a criterion of interest, which in the context of QST would be the likelihood of the measurement data acquired.

In this work, we apply variational approaches to perform tomography of \teal{complex multiqubit} quantum states produced in a $N=20$-qubit trapped-ion quantum simulator, which corresponds to a Hilbert space of dimension $2^{20}\approx10^6$. The system evolves under the action of an engineered many-body Hamiltonian \cite{qubits20}  \teal{generating highly entangled states. Efficiently characterizing these states poses a current challenge due to their non-symmetrical nature, which distinguishes them from highly symmetric states such as Greenberger-Horne-Zeilinger (GHZ) or W states, for which effective tomographic methods are available.} Nevertheless, successful state reconstruction is achieved in spite of very limited data (statistics collected over 27 bases with 1000 measurements in each basis).

In our present work, we compare the performance of three methods that belong to the following two classes: the matrix-product states (MPS) ansatz and two neural-network quantum state (NNQS) ans\"atze, a restricted Boltzmann machine and an autoregressive neural network. Both classes have their advantages and shortcomings. The MPS ansatz is easier to train, but works best on chains of qubits with short-range one-dimensional correlations (more generally, area-law entanglement scaling) and limited degree of entanglement \cite{orus2019tensor,gomez2022reconstructing}. In fact, a version of the MPS ansatz has been shown to fail on the specific system studied here \cite{qubits20}. On the other hand, NNQSs are more expressive \cite{sharir2022neural} and have been shown to work well on states with volume law entanglement scaling~\cite{deng2017quantum,levine2019quantum,hibat2020recurrent}. However, they are more difficult to train: they require sampling and, furthermore, transforming the sampled state into the measurement basis results in emergence of an exponential number of terms. Therefore, there is no obvious answer to the question of which ansatz is more suitable in our case. 

Our results unequivocally demonstrate that MPS surpasses both neural networks in terms of likelihood for the experimental data and in terms of fidelity for simulated ones. We believe this to be a consequence of the one-dimensional geometry and short-range character of interactions in our trapped-ion simulator. To illustrate this point, at the end of the paper we provide an example of a volume-law scaling state whose tomography can be successfully achieved with NNQS, but fails with MPS. 

\subsection*{Setup and measurement bases}
We study quantum states of a 20-qubit trapped-ion system, which is described in Ref.~\cite{qubits20}, at eight different time points during its evolution. We briefly recap the description of the experiment. The setup is based on the controllable one-dimensional array of $^{40} \mathrm{Ca}^{+}$ ions, which are confined in a linear Pauli trap with axial (radial) center-of-mass vibrational frequency of 220 kHz (2.712 MHz) \cite{hempel2014digital}. 
More information about the state preparation and measurement process can be found in Ref.~\cite{qubits20}.  
The computational qubit basis consists of the states $|S_{J=1/2;m_j=1/2}\rangle$ and  $|D_{J=5/2;m_j=5/2}\rangle$, connected by an electric quadrupole transition at
729 nm. The leakage rates outside this qubit manifold are on a scale of \teal{1 Hz} \cite{schindler2013quantum}, i.e.~negligible in comparison with the time scales we are working with here.

The qubit interactions under the influence of laser-induced forces are well described by the XY model in a dominant transverse field $B$, with the following Hamiltonian: 
\begin{equation}\label{Ham}
	\hat{\mathcal{H}}=\hbar \sum_{i<j} J_{i j}\left(\sigma_{i}^{+} \sigma_{j}^{-}+\sigma_{i}^{-} \sigma_{j}^{+}\right)+\hbar B \sum_{j} \sigma_{j}^{z}
\end{equation}
where $J_{i j}$ is the $N \times N$ qubit-qubit coupling matrix, 
$\sigma_{i}^{+}\left(\sigma_{i}^{-}\right)$ is the qubit raising (lowering) operator for qubit $i$, and $\sigma_{j}^{z}$ is the Pauli $Z$ matrix for qubit $j$. The coupling strengths fall approximately according to a power law $J_{i j} \propto 1 /|i-j|^{\alpha}$ with the distance $|i-j|,$ where $\alpha\approx1.1$.
The 20-qubit register is initiated to a N{\`e}el-ordered product state $|\Psi(0)\rangle=|1, 0,1, \ldots\rangle$. It is then expected to evolve according to the Schr\"odinger equation
\begin{equation}\label{SchEq}
	|\Psi_{\rm th}(t)\rangle=e^{-i\hat{\mathcal{H}}t}\ket{\Psi(0)}.
\end{equation}
The interaction initially gives rise to entanglement among neighboring qubits, which then spreads over the entire quantum register \cite{lanyon2017}. This evolution is affected by experimental imperfections and decoherence. However, the chosen experimental regime creates a decoherence-free subspace during the evolution \cite{joshi2022}. Residual decoherence from imperfections in the laser-ion interaction is much slower than the evolution time scales studied here, hence it is fair to assume that the state purity does not degrade on these time scales.

The set of measurement bases is defined by the observables of the form $\sigma_a\otimes\sigma_b\otimes\sigma_c\otimes\sigma_a\otimes...\otimes\sigma_c$ of periodicity $k=3$ and length 20, with $a,b,c$ being all 27 possible sequences of $X,Y,Z$. This choice of the basis set was made in Refs.~\cite{qubits20, lanyon2017}, motivated by its sufficiency to observe genuine multi-partite entanglement in groups of up to five qubits. \teal{By increasing the periodicity of the basis pattern, one could capture the entanglement in longer groups. This would become necessary for longer evolution times under the nearest-neighbor Hamiltonian since the correlation length grows with time. Knowing the speed at which the correlations propagate through the system, one can theoretically predict the correlation length and hence the period of the measurement basis sufficient for reliable characterization of the quantum state~\cite{schachenmayer2013entanglement}.}

In this work, we use the same measurement data set as in Ref.~\cite{qubits20}, which includes 1000 measurements per basis (limited by the stability of the experimental setup). As we show here, this set is sufficient for QST of the 20-qubit system.

\subsection*{Variational many-body state tomography}\label{sec:Variational_Approach}
The complete description of any pure quantum system constituting $N$ qubits is given by $|\Psi\rangle = \sum_{\boldsymbol{s}} \Psi(\boldsymbol{s})|\boldsymbol{s}\rangle$, 
where $\Psi(\boldsymbol{s})$ represents the probability amplitude corresponding to the computational basis element $\boldsymbol{s}\equiv\left(s_1,s_2,\dots,s_N\right)$, where $s_i= 0$ or $1$, which we refer to as the bit configuration. 
Our goal is to reconstruct a useful many-body pure quantum state $|\Psi_W\rangle$, which maximizes the probability (likelihood) of having acquired the given measurement data. For numerical stability of gradient descent, it is more convenient to minimize the negative logarithm of that probability, which we hereafter refer to as the loss function:
\begin{equation}\label{LL}
	\Xi = -\sum_{p=1}^K\sum_{m=1}^{M_p}\ln\left(\abs{\langle\psi_m^p|\Psi_W\rangle}^2+\varepsilon \right),
\end{equation}
where $K=27$ is the number of measurement bases, $M_p=1000$ is the number of measurements per basis, $\ket{\psi_m^{p}}$ is the outcome of the $m$th projection measurement in $p$th basis. A small term $\varepsilon=10^{-10}$ is added to improve the reconstruction stability, as discussed in the Supplementary. The subscript $W$ in $\ket{\Psi_W}$ defines the set of parameters of the variational ansatz used. In this study, the size of $W$ was chosen to be approximately 4000 in each ansatz.
At first, parameters $W$ are randomly initialized, and then the variational parameters $W$ are adjusted (trained) to minimize the negative log-likelihood~(\ref{LL}) by gradient descent. 

To prevent overfitting, we split the measurement data into two sets, training and test, in the ratio 4:1 \cite{joseph2022optimal}. Specifically, we considered the first 800 measurements in each basis to be the training data and the remaining 200 measurements to be the test data. The loss function is optimized on the training set, and the test set is used to validate the performance of different methods. The training process is stopped as soon as the loss on the test set reaches its minimum.

An essential component of multi-qubit tomography is the transformation (rotation) of the variational state $|\Psi_W\rangle$ into the measurement basis, required to compute the probability of each measurement outcome. We can express this mathematically as 
\begin{equation}\label{overlap}
    \langle\psi_m^p|\Psi_W\rangle=\langle \mathbf{j}_m^p|\hat U_p|\Psi_W\rangle,
\end{equation}
where $\ket{\mathbf{j}_m^p}$ is the state in the computational basis defined by the bit string yielded by the measurement, and $\hat U_p$ is the rotation between the computational basis and the $p$th measurement basis. The problem is that an arbitrary rotation can in principle lead to a vector with exponentially many terms, even if the initial state contains only a polynomial number of nonzero components. We address this problem differently for the MPS and NNQS ans\"atze.

\subsection*{Matrix product states}
A matrix product state is a tensor network, which represents each amplitude in a quantum state as a product of matrices,
\begin{equation}\label{MPS}
	|\Psi_{\rm mps}\rangle = \sum_{\boldsymbol{s}} \left[ A_{s_1}^{1}A_{s_2}^{2} \dotsb A_{s_N}^{N} \right]\ket{s_1\,s_2\,\dots \, s_N},
\end{equation}
where $A_{s_i = 0}^{i}$ and $A_{s_i = 1}^{i}$ are the complex matrices corresponding to the {\it i}th qubit. The sizes of these matrices are hyperparameters of the MPS ansatz, which are referred to as bond dimensions.  
In our work, the matrices $A_{s_1}^{1}$ and $( A_{s_N}^{N})^\teal{\top}$ are of size $1\times 2$, $A_{s_2}^{2}$ and $(A_{s_{N-1}}^{N-1})^\teal{\top}$ are of size $2\times 4$, $A_{s_3}^{3}$ and $(A_{s_{N-2}}^{N-2})^\teal{\top}$ are of size $4\times 8$, $A_{s_4}^{4}$ and $(A_{s_{N-3}}^{N-3})^\teal{\top}$ are of size $8\times 10$,  and all others are of size $10\times 10$ \cite{schollwock2011density}. The maximal bond size of 10 is chosen as a hyperparameter.

Our ansatz differs from a previous study on the applicability of MPSs for QST of trapped-ion qubits~\cite{lanyon2017}. In that work, Lanyon {\it et al.} implemented full mixed state tomography of each neighboring qubit triplet in ion chains of sizes up to $N=14$, and then found a pure-state MPS that approximates the combined set of these three-qubit density matrices. This approach was motivated by the existence of a lower bound on the fidelity between the reconstruction and the true state~\cite{cramer2010efficient}. For 20-ion chains, this bound was found to be low, leading to the conclusion that ``MPS tomography failed to produce a useful pure-state description in our present 20-qubit system"~\cite{qubits20}. 
However, this conclusion only applies to the specific QST strategy of Ref.~\cite{lanyon2017}. This strategy, while allowing for full characterization of each triplet of neighboring qubits, and hence estimating the lower bound on fidelity, does not account for the information on long-range correlations, which can be acquired through simultaneous measurements on all qubits. Our present ansatz addresses this shortcoming.

A significant benefit of the MPS approach to QST is the simplicity of rotating the state to the measurement basis (provided that the basis is local, i.e.~is obtained from the computational basis by means of single-qubit rotations, as is in our case). The matrices $(A^i_{s_i})$ and  $(A^i_{s_i})'$ defining the states $|\Psi_{\rm mps}\rangle$ and $\hat U_p|\Psi_{\rm mps}\rangle$ are related according to  
\begin{equation}
    (A^i_{s_i})' = \sum_{s'_i = 0}^1\hat{V^i}_{s_i s'_i} A^i_{s'_i}\;,
\end{equation}
where $\hat{V}^i=e^{i\frac\pi4\hat\sigma^i}$ is the $2\times2$ matrix defining the SU(2) rotation to the measurement basis of the $i$th qubit, which corresponds to the observable $\hat\sigma^i$ \cite{orus2019tensor} and the measurement basis index $p$ is omitted for brevity. Consequently, the likelihood \eqref{LL} can be directly expressed in terms of the ansatz parameters and the bit strings with the measurement results.

\begin{figure*}[t!]
    \centering
    \includegraphics[width=0.85\textwidth]{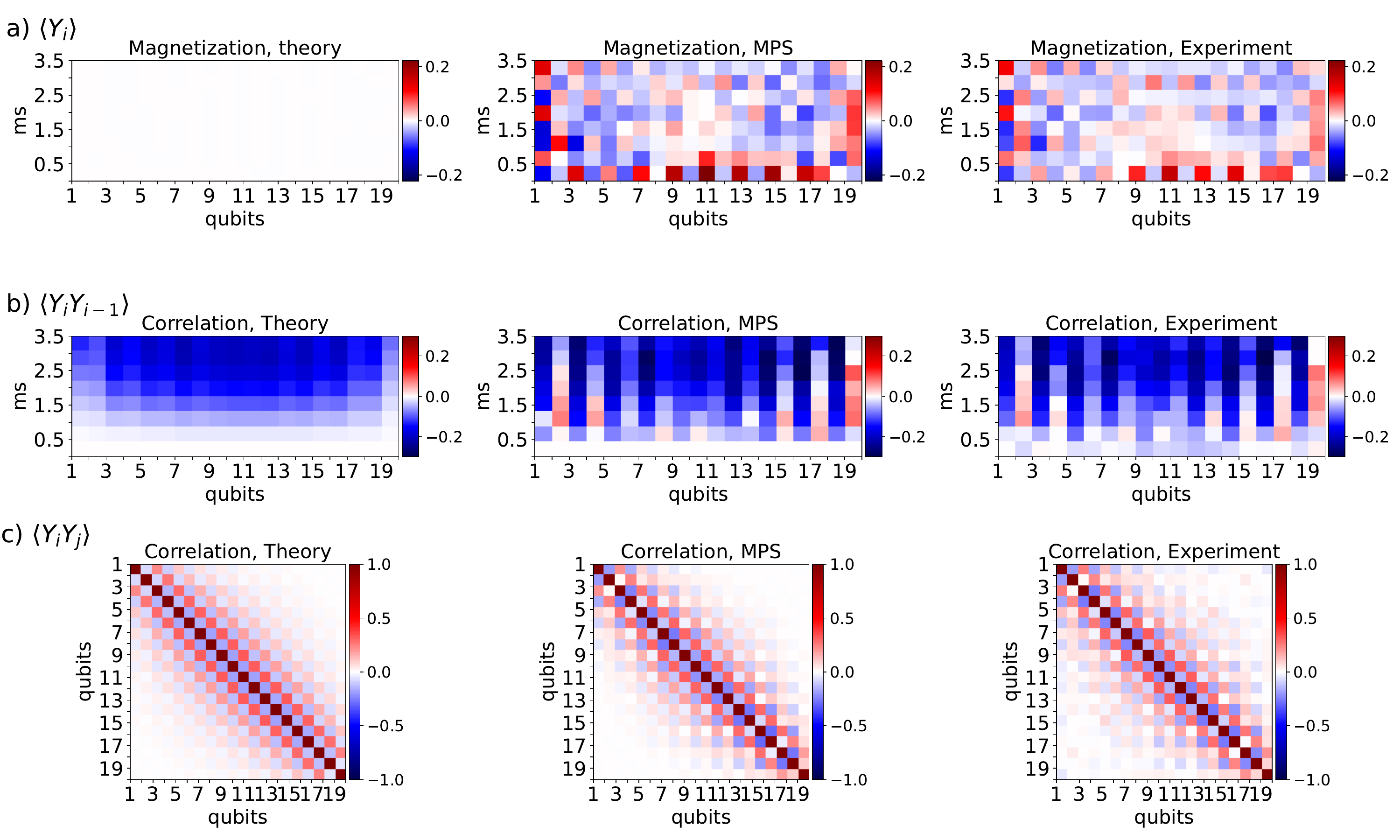}
    \caption{Estimation of observables in the experimental and reconstructed states. For the observable $\hat\sigma_Y$, a) single-qubit expectation values; b) correlations for neighboring qubits; c) full pairwise correlations are shown. The vertical axis in (a) and (b) enumerates the eight moments in time for which the measurements were made; the data in (c) correspond to $t=3$ ms. The left columns show the theoretically expected evolution \eqref{SchEq}, the central column the experimental MPS reconstruction, and the right column is obtained directly from the experimental data. The empty rectangle in the left column of (a) means that the theoretically expected magnetization is constantly zero.}
    \label{fig:2_qubit_correlations}
\end{figure*}

\subsection*{Neural networks}
Here we utilize two different NNQS ans\"atze, specifically, a restricted Boltzmann machine (RBM) and an autoregressive neural network (ARN). Both neural networks have two functions: inference and sampling. When used for inference, the neural network outputs the probability amplitude $\Psi(\boldsymbol s)$ for an input bit configuration $\boldsymbol s$. When used for sampling, the neural network outputs bit configuration samples of the probability distribution $|\Psi(\boldsymbol s)|^2$. The network has to be run multiple times to produce a large ensemble of samples, which approximates the state vector $|\Psi\rangle = \sum_{\boldsymbol{s}} \Psi(\boldsymbol{s})|\boldsymbol{s}\rangle$. The loss function \eqref{LL} can then be computed from this vector and the neural network parameters can be adjusted to minimize it. The two networks are different in construction and functionality; here we present a brief literature overview for both types of NNQS and the details of our implementation thereof. A more introductory discussion is provided in the Supplementary. 

RBM has been used for a large variety of quantum state analysis problems~\cite{carleo2017solving,glasser2018neural,jonsson2018neural,deng2017quantum, deng2017machine, nomura2017restricted,nomura2017restricted,luo2019backflow, han2019solving,saito2017solving}, in particular, for QST in both discrete- and continuous-variable settings~\cite{Torlai2018, torlai2018latent, tiunov2019experimental, 
torlai2019integrating}. 
However, RBMs have their limitations: 
there is a known class of physically interesting quantum states that cannot be efficiently approximated by this ansatz~\cite{gao2017efficient}.
In addition, the probability amplitudes output by the RBM are unnormalized, requiring a special procedure for estimating the partition function during training. 

In contrast, the ARN approach gives normalized amplitudes explicitly~\cite{wu2019solving}, leading to an effective and unbiased sampling procedure.
The ARN method has been recently used for solving quantum many-body problems~\cite{sharir2020deep}, in particular, in quantum chemistry~\cite{barrett2022autoregressive}. A further advantage of ARN is its ability to generate non-redundant samples only, which greatly reduces the required computational resources~\cite{barrett2022autoregressive}.

We use two RBMs of identical architectures for the amplitudes and phases of qubit configurations, as proposed in the original work by Carleo and Troyer \cite{carleo2017solving}. Each RBM has 100 hidden units, with the sampling implemented via one-step persistent contrastive divergence~\cite{carreira2005contrastive,tieleman2008training}, i.e.~such that the samples from each previous iteration are reused in the next one. Our ARN consists of two hidden layers with the size of each layer equal to $N=20$ as required by the autoregressive architecture (see Supplementary).  To enhance the network's expressibility, each unit in the hidden layers is a cluster of three numbers \emph{\`{a} la} Refs.~\cite{made,nade, papamakarios2017masked}. Each unit of the output layer is a cluster of two numbers corresponding to the conditional amplitude and phase of the spin configuration. The ADAM optimizer \cite{Adam2015} with the learning rate $\gamma=0.005$ is used for reconstructions. Convergence {of the ARN ansatz} is typically achieved in less than 1000 epochs, with a runtime of around 30 minutes, using a single NVIDIA V100 Volta-based card, and a memory requirement of no more than 4 GB. {Under the same conditions, 10000 epochs are needed to reach the same level of convergence of the RBM ansatz.}

To address the qubit rotation issue in the NNQS ansatz, we again use the fact that all measurement bases are local. Denoting $\mathbf{j}_m^p =j_1 \ldots j_N$ and $\hat U_p=\hat{V}_1\otimes \ldots \otimes\hat{V}_N$, and recalling that $|\Psi\rangle = \sum_{\boldsymbol{s}} \Psi(\boldsymbol{s})|\boldsymbol{s}\rangle$ with $\ket{\boldsymbol{s}}=\ket{s_1 \ldots s_N}$ (where $\ket{\boldsymbol{s}}$ in this case are the sampled bit configurations), the overlap \eqref{overlap} can be written as 
\begin{equation}\label{eq:matrix_element2}
\begin{aligned}
&	\langle \mathbf{j}_m^p|\hat{U}_p| \Psi_W \rangle = \\  
& \sum_{\textbf{s}} \Psi(\textbf{s}) \langle j_1 \ldots j_{N} | \hat{V}_1 \otimes \ldots \otimes \hat{V}_{N} | s_1 \ldots s_{N} \rangle = \\
& \sum_{\textbf{s}} \Psi(\textbf{s}) \langle j_1|\hat{V}_1| s_1\rangle\times\ldots\times\langle j_N|\hat{V}_N| s_N\rangle.
\end{aligned}
\end{equation}
We see that the overlap is a product of $N$ complex numbers, which must be calculated $K \times M_p\times |\{\boldsymbol{s}\}|$ times, where $K$ represents the number of measurement bases and $|\{\boldsymbol{s}\}|$ is the number of unique samples produced by NNQS. This calculation becomes tractable provided that the NNQS with the given $|\{\boldsymbol{s}\}|$ is expressive enough to represent the optimal state of interest.

\subsection*{Results}
The central column of Fig.~\ref{fig:2_qubit_correlations} shows the qubit values and their correlations obtained from the states  $|\Psi_{W}(t)\rangle$ reconstructed by the MPS corresponding to eight moments in time ranging from $t=0.0$ to $3.5$ ms with a $0.5$ ms interval, for which the measurements are performed. 
For comparison, we show the results corresponding to the theoretically expected evolution according to the Hamiltonian of Eq.~\eqref{SchEq}
\cite{qubits20} (left column) and the direct measurement results (right column). We see that the agreement between the reconstructed state and the data is much better than that between either of the two and the theoretical model. \teal{In the supplementary, we estimate the statistical uncertainty of our reconstruction and find that the deviation from the theoretical state is statistically significant~\cite{bar2001estimation}.}

\begin{figure}[tbh!]
 \includegraphics[width=0.45\textwidth]{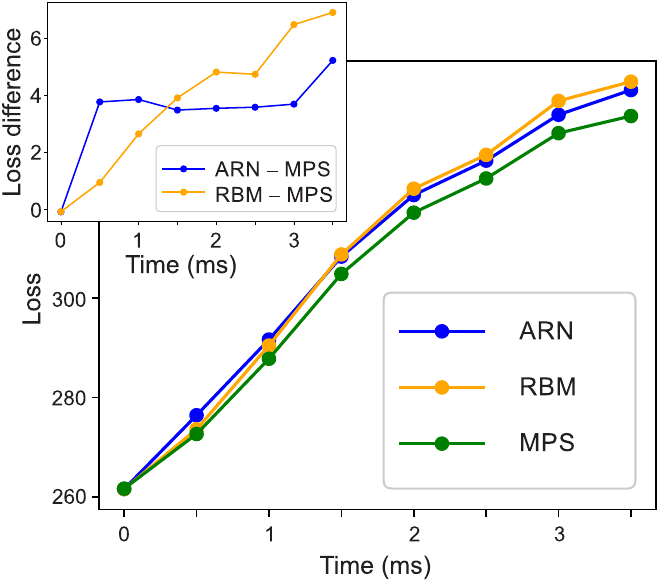}
 \caption{\teal{Comparison of the loss (negative log-likelihood) \textcolor{blue}{on the test data} for the  experimentally reconstructed states. The data shown represent different evolution times in the experiment and different reconstruction methods. For each point, the number of samples drawn from the NNQS ans\"atze is 300000. The inset shows the difference of the three methods.}}
 \label{fig:loss_allstates}
\end{figure}

Figure \ref{fig:loss_allstates} shows the comparison between the loss Eq.~\eqref{LL} on the test sets for the states reconstructed via MPS and the two NNQS ans\"atze. We observe that the ARN and RBM exhibit comparable performance, while the MPS generally performs better. 

\begin{figure}[bt!]
\includegraphics[width=0.48\textwidth]{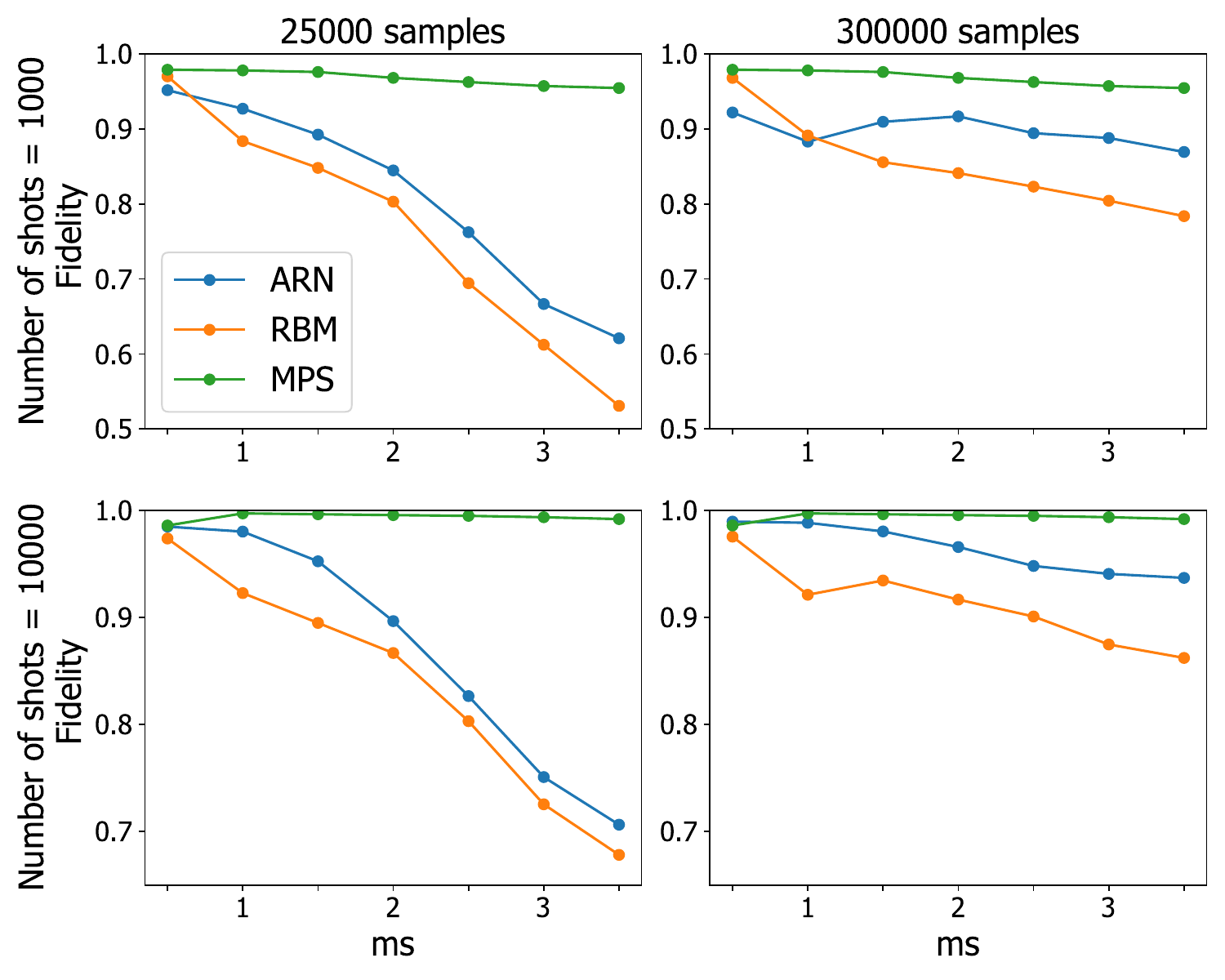}
\caption{Fidelities of the states reconstructed from the simulated data generated from the theoretical states \eqref{SchEq} with respect to these states. The performance of the NNQS ans\"atze improves with the number of samples drawn from the NN during training (left vs right plot). For all methods, the performance improves with the number of experimental shots (top vs bottom plots).
}
\label{fig:fidelity_auto_vs_feed}
\end{figure}

\begin{figure*}[htb!]
\includegraphics[width=0.75\textwidth]{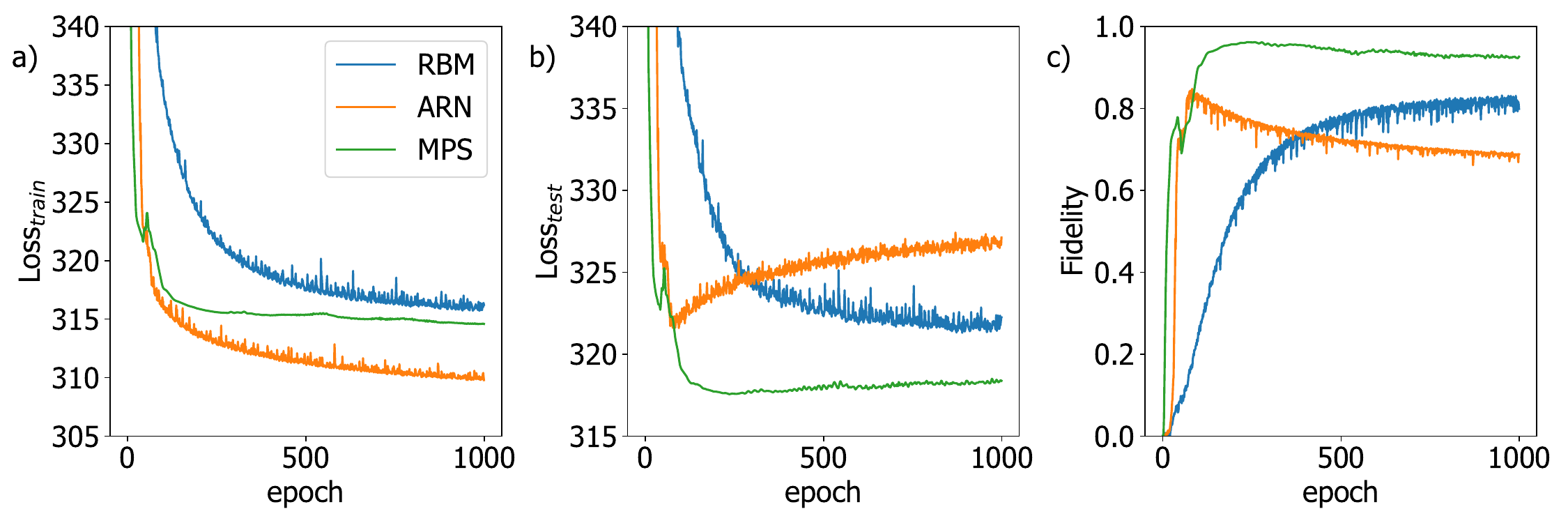}
    \caption{Learning curves (the progress in the loss function and fidelity in the process of learning) for different ans\"atze reconstructing $\ket{\Psi_{\rm th}(t=2.5\textrm{ ms})}$ from simulated data with 1000 measurements per basis. Losses on the training (a) and test (b) sets are shown, as well as the fidelity between the reconstructed and ground truth states (c). The number of samples drawn from the NNs during training is $300000$. The observed kinks are a consequence of automatic adjustment of the learning rate by the ADAM optimizer.}
    \label{fig:Overfitting_analysis}
\end{figure*}

The above results do not yet present compelling evidence of the reconstruction quality, as the states that generated the measurement data---the \emph{ground truth} for the machine learning problem---are not known. Hence, in addition to QST based on experimental data, we perform a series of independent numerical experiments in which we reconstruct a state from simulated measurement data generated from the theoretically expected state according to Eq.~\eqref{SchEq} in the given 27 measurement bases. This allows us to quantify the fidelity of the reconstructed states with the true underlying state, as shown in Fig.~\ref{fig:fidelity_auto_vs_feed}. We observe again that the reconstruction fidelity $|\langle \Psi_{\rm th}(t) |\Psi_{W}(t)\rangle|^{2}$ obtained with the MPS consistently surpasses the neural-network methods for all evolution times. 
Importantly, Fig.~\ref{fig:fidelity_auto_vs_feed} shows that despite the low number of experimental shots, the reconstruction fidelity of the NNQS ans\"atze improves significantly with an increased number of samples drawn from the neural network for each training step.
For all methods, the fidelity also decreases with increasing time $t$ as the evolving state becomes more complex.\par

As shown above, MPS outperforms NNQSs in terms of likelihood for the experimental data and in terms of fidelity for the simulated one. One can suspect two possible reasons for that: insufficient expressive power of NNQS methods or failure of training. To resolve this dichotomy, we test for insufficient expressivity by training the NNQSs to directly maximize the fidelity with the theoretical ground truth states rather than the likelihood of the simulated measurement data. As a result, we obtained fidelities exceeding 99\% for all 8 states of the time evolution. We note that the training on simulated data is merely for developing a better understanding of the behavior of the NNQSs and was not used when reconstructing experimentally measured quantum states. These results indicate that the neural network ans\"atze are in principle sufficiently expressive, meaning that there exists a parameter set for which the NNQS is able to approximate the ground truth state to within $1\%$ fidelity, at least for the simulated data. This is consistent with the conclusion of Refs.~\cite{deng2017quantum,sharir2021neural} that neural networks are able to represent a wide range of quantum states. Moreover, these states demonstrate lower loss on the simulated datasets than the NNQSs trained to maximize the likelihood. These observations lead us to conclude that the poorer performance of the NNQS is likely attributed to their challenging training. We assume that this is due to a complex loss function landscape in the NNQS parameter space. The improvement of NNQS fidelity with more data could then be due to this landscape getting smoother with fewer local optima, so the neural network becomes easier to train.

Figure~\ref{fig:Overfitting_analysis} shows the analysis of overfitting in our training process. The loss on the training set monotonically decreases with training [Fig.~\ref{fig:Overfitting_analysis}(a)] while the loss on the test set, previously unseen by the reconstruction algorithm, starts to increase at some point during training [Fig.~\ref{fig:Overfitting_analysis}(b)]. This is because the measurement data are samples of a statistical distribution associated with the quantum state. Maximizing the likelihood for a specific set of samples may result in a reconstructed state that is overfitted to this sample \cite{mogilevtsev2013cross,steffens2017experimentally}. Validating the reconstructed state on the test dataset enables us to control this overfitting. 

As seen in Fig.~\ref{fig:Overfitting_analysis}(b), the lowest loss on the test set corresponds to the maximum of the fidelity curve [Fig.~\ref{fig:Overfitting_analysis}(c)]. This indicates that monitoring the loss on the test set is a good independent indicator for deciding when to stop the training to prevent overfitting, particularly in the case of the ARN, which shows a pronounced dip (peak) in the loss on the test set (fidelity with the true states) in Fig.~\ref{fig:Overfitting_analysis}.

We observe that MPS trains significantly faster than NNQSs.

\begin{figure*}[tbh!]
 \includegraphics[width=\textwidth]{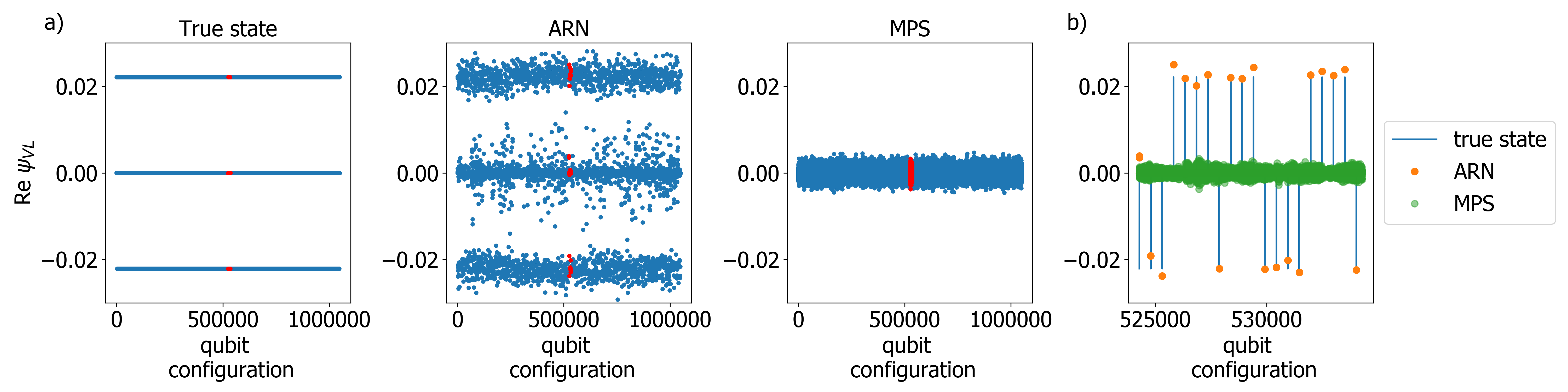}
 \caption{QST reconstruction of the volume law state of Eq.~\eqref{eq:VL_state} from simulated data. The horizontal axis indexes bit configurations from $1$ to $2^{20}$, the vertical axis shows the real part of the corresponding amplitude. a) Amplitudes for the entire set of possible spin configurations. b) Data for a small interval of bit strings showing consistency between the true state and reconstructions. The corresponding points in (a) are shown in red.\label{fig:VL_state}}
\end{figure*}

\subsection*{MPS is not always better}
The above results demonstrate the advantage of the MPS ansatz, manifesting  in both faster convergence and higher  reconstruction quality. This advantage is a consequence of our system being a 1D qubit chain with area law entanglement scaling dynamics, which falls within the scope of this ansatz. One should keep in mind, however, that this scope is quite limited. For example, its performance can drop significantly if the state's entanglement scales according to the volume law, in which case exponential bond dimension scaling is required \cite{orus2019tensor}. To illustrate this point, we apply QST to a 1D state with volume law scaling discussed in Ref.~\cite{deng2017quantum}:
\begin{equation}\label{eq:VL_state}
    \ket{\Psi_{\textit{VL}}}=\frac1{2^{N/4}}\sum\limits_{\boldsymbol{s}} (-1)^s \ket{\boldsymbol{s}\boldsymbol{s}},
\end{equation}
where the summation is over all possible bit strings $\boldsymbol{s}$ of length $N/2$, and the allowed configurations of the $N$ qubits are such that the first half of the string is identical to the second.
 We artificially generate 1000 outcomes per basis for the same set of 27 measurement settings and reconstruct the state via MPS and ARN, resulting in fidelities 0.000081\% and 88.1\%, respectively. The amplitudes of the underlying ideal state of Eq.~\eqref{eq:VL_state} and of the reconstructed states are depicted in Fig.~\ref{fig:VL_state}. Hence, the results presented in this work ought not to be viewed as an argument against NNQS's ability in representing complex, multidimensional entangled states.

\subsection*{Summary}
We performed quantum-state reconstruction of a 20-qubit state (in the pure-state approximation), which, to our knowledge, is the largest experimental system to which QST has been applied \cite{riofrio2017experimental,lanyon2017,torlai2019integrating}. The state that was reconstructed is the result of quench dynamics under an XY Hamiltonian. The XY model is an archetypal model for the study of quantum many-body phenomena, including phase transitions, quasi particle dynamics, and entanglement propagation. The XY model is known
to be a universal simulator of Hamiltonians in 2D  and thus has broad significance for the field of quantum simulation~\cite{cubitt2018universal}.

Here, we focused on the time evolution of many-body quantum systems following a quench, which is a challenging problem, even in one dimension \cite{cirac2012goals}. By examining the growth of entanglement in quantum dynamics, we are able to not only gain valuable insights into the underlying microscopic processes but also explore the complexity of quantum states. This information is directly connected to the efficiency of simulations conducted on classical computers \cite{schachenmayer2013entanglement}. As using quantum quench dynamics one can generate states that contain substantial amount of entanglement, they can be used as a test bed to understand the underlying dynamics of challenging quantum many-body problems from various domains including condensed matter physics, quantum statistical mechanics, high-energy physics, atomic physics and quantum chemistry.

The number of variational parameters ($\sim 4000$) in all methods is very low compared to the Hilbert space dimension ($2^{20}\approx 10^6$). More importantly, the number of measurement bases (27) is far below the QST quorum ($\ge 2^{20}+1$), and the number of measurements in each basis is far below the dimension of the Hilbert space. Nonetheless, the reconstructed states reproduce the true state of the system very well. This demonstrates that under the right circumstances, state reconstruction remains possible even from highly incomplete data sets.

Among the methods used, RBM and ARN show comparable reconstruction quality, while MPS outperforms both NNQSs in terms of both the experimental data likelihood and fidelity with ideal simulated states. Moreover, the MPS converges faster and generalizes better (overfits less) on average than the neural networks.

Inquiring into the reasons for this disparity in performance, we found that it is not the expressive power of NNQSs that limits their fidelity, but the complexity of their training. One can identify two issues here. 
The first one is the state's transformation (rotation) into different measurement bases. A rotated MPS can be readily expressed as another MPS with a different parameter set, whereas rotating an NNQS is much more complicated. Second, the data likelihood can be directly expressed in terms of MPS parameters, whereas NNQSs require an expensive and imprecise sampling procedure. Both these features complicate the training of neural-network ans\"atze for quantum state representation.   

On the other hand, the superiority of MPS is linked to the physical nature of quantum states that are produced by the 1D quantum simulator, in particular, the dominance of nearest-neighbor correlations. Such states are known to be well described by MPS. The class of states for which NNQS may show superior performance in QST is states with volume-law entanglement scaling, as shown in Fig.~\ref{fig:VL_state}. 

A natural next step would be to extend our method to mixed states. A challenge associated therewith is the quadratic increase in the number of parameters in a density matrix as compared to a pure state vector. This can be mitigated by combining machine learning methods with quantum-state tomography by compressed sensing \cite{Gross_2010}. An alternative is to directly apply the MPS formalism to mixed states \cite{verstraete2004matrix}.

The code and data used in this paper are available on Zenodo \cite{our_code}.

\subsection*{Acknowledgments}
The scientific work was completed between 2019 and 2021. M.K.K. acknowledges the support provided by University of Calgary and Compute Canada (\url{www.computecanada.ca}) for High Performance Computing facilities.
A.K.F. thanks the support of the RSF grant (19-71-10092).
The Innsbruck team acknowledges support by the Austrian Science Fund (FWF), through the SFB BeyondC (FWF Project No.~F7109) and the Institut f\"ur Quanteninformation GmbH. We also acknowledge funding from the EU H2020-FETFLAG-2018-03 under Grant Agreement No.~820495 and the European Union’s Horizon 2020 research and innovation program under Grant Agreement No.~840450.
 
\bibliography{references.bib}

\end{document}


\preprint{APS/123-QED}

\title{Supplementary information: Reconstructing complex states of a 20-qubit quantum simulator}

\author{Murali K. Kurmapu}
\affiliation{University of Calgary, Calgary AB T2N 1N4, Canada}
\affiliation{Department of Physics, University of Oxford, Oxford OX1 3PG, UK}

\author{V.V. Tiunova}
\affiliation{Russian Quantum Center, Skolkovo, Moscow 143025, Russia}

\author{E.S. Tiunov}
\affiliation{Quantum Research Centre, Technology Innovation Institute, Abu Dhabi, UAE}

\author{Martin Ringbauer}
\affiliation{Institut f\"{u}r Experimentalphysik, Universit\"{a}t Innsbruck, Technikerstrasse 25, 6020 Innsbruck, Austria}

\author{Christine Maier}
\affiliation{Alpine Quantum Technologies GmbH, 6020 Innsbruck, Austria}

\author{Rainer Blatt}
\affiliation{Institut f\"{u}r Experimentalphysik, Universit\"{a}t Innsbruck, Technikerstrasse 25, 6020 Innsbruck, Austria}
\affiliation{Alpine Quantum Technologies GmbH, 6020 Innsbruck, Austria}
\affiliation{Institut f\"{u}r Quantenoptik und Quanteninformation, \"{O}sterreichische Akademie der  Wissenschaften, Otto-Hittmair-Platz 1, 6020 Innsbruck, Austria}

\author{Thomas Monz}
\affiliation{Institut f\"{u}r Experimentalphysik, Universit\"{a}t Innsbruck, Technikerstrasse 25, 6020 Innsbruck, Austria}
\affiliation{Alpine Quantum Technologies GmbH, 6020 Innsbruck, Austria}

\author{Aleksey K. Fedorov}
\affiliation{Russian Quantum Center, Skolkovo, Moscow 143025, Russia}
\affiliation{National University of Science and Technology ``MISIS'', 119049 Moscow, Russia}

\author{A.I. Lvovsky}

\affiliation{Department of Physics, University of Oxford, Oxford OX1 3PG, UK}

\maketitle

\section{Restricted Boltzmann machine}
The restricted Boltzmann machine (RBM) is a two-layer generative neural network. Neurons of both layers take binary values. The first layer, containing the bit string configuration of interest, is called visible ($\boldsymbol{s}$) and the second one is called hidden ($\boldsymbol{h}$) [Fig.~\ref{fig:model}(a)]. A joint configuration of the two layers defines a joint Boltzmann-like probability 
\begin{equation*}
    p(\boldsymbol{s},\boldsymbol{h}) = \frac{1}{Z} \exp(\boldsymbol{s}W\boldsymbol{h} + \boldsymbol{a}\cdot\boldsymbol{s} + \boldsymbol{b}\cdot\boldsymbol{h}),
\end{equation*}
where $W$ is the weight matrix, $\boldsymbol{a},\boldsymbol{b}$ the biases of the hidden and a visible layers respectively, and $Z$ is a partition function. The usual goal of the RBM training is to find the marginal distribution $p(\boldsymbol{s})$
\begin{equation}\label{prvRBM}
    p(\boldsymbol{s}) = \sum_{\boldsymbol{h}} p(\boldsymbol{s},\boldsymbol{h}),
\end{equation}
 which is optimal with respect to a certain criterion. The summation in the above equation  is over all possible configurations of a hidden layer. 
 
We follow the variational approach to obtain the RBM representation of an arbitrary quantum state. A multi-qubit state can be represented as follows:
\begin{equation}\label{psiRBM}
    \ket{\Psi_W} = \sum_{\boldsymbol{s}} \sqrt{p(\boldsymbol{s}}) e^{i \phi(\boldsymbol{s})} \ket{\boldsymbol{s}},
\end{equation}
where $\boldsymbol{s}$ the qubit configuration in the computational basis.  To represent the amplitudes $\sqrt{p(\boldsymbol{s})}$ and phases $\phi(\boldsymbol{s})$, we use two RBMs in the following manner:
\begin{subequations}\label{RBMeq}
\begin{eqnarray}
    p(\boldsymbol{s}) = \frac{1}{Z} \sum_{\boldsymbol{h}} \exp(\boldsymbol{s}W_{\lambda}\boldsymbol{h} + \boldsymbol{a}_{\lambda}\cdot\boldsymbol{s} + \boldsymbol{b}_{\lambda}\cdot\boldsymbol{h})\\
    \phi(\boldsymbol{s}) = \log\left[\sum_{\boldsymbol{h}} \exp(\boldsymbol{s}W_{\mu}\boldsymbol{h} + \boldsymbol{a}_{\mu}\cdot\boldsymbol{s} + \boldsymbol{b}_{\mu}\cdot\boldsymbol{h})\right],
\end{eqnarray}
\end{subequations}
where the subscripts $\lambda$ and $\mu$ denote the parameters of the  amplitude and phase RBMs respectively~\cite{Torlai2018}. These expressions can be simplified since there are connections only between visible and hidden layers:
\begin{equation*}
   \sum_{\boldsymbol{h}} \exp(\boldsymbol{s}W_{\lambda,\mu}\boldsymbol{h} + \boldsymbol{a}_{\lambda,\mu}\cdot\boldsymbol{s} + \boldsymbol{b}_{\lambda,\mu}\cdot\boldsymbol{h}) = 
   \exp\left[  \boldsymbol{a}_{\lambda,\mu} \cdot\boldsymbol{s} + 
	\sum_i \log\left( 1 + e^{\left(\boldsymbol{s} W_{\lambda,\mu} + \boldsymbol{b}_{\lambda,\mu}\right)_i} \right) \right]. 
\end{equation*}
Equations \eqref{RBMeq} define a variational Ansatz parameterized by the two sets of weights and biases. These parameters must be optimized with respect to the log-likelihood functional.


Sampling in an RBM is realized using  the algorithm known as contrastive divergence (CD)~\cite{carreira2005contrastive}.  One step of the CD algorithm consists of two actions. The first one is sampling configurations of the hidden layer given the configurations of the visible layer according to the conditional probabilities $p(\boldsymbol{h}|\boldsymbol{s})$. The second operation is reverse:  we sample the visible layer configurations according to the conditional probabilities $p(\boldsymbol{s}|\boldsymbol{h})$. 
\begin{eqnarray*}
    p(\boldsymbol{h} = \boldsymbol{0}|\boldsymbol{s}) = \sigma(\boldsymbol{s} W + \boldsymbol{b});\\
    p(\boldsymbol{s} = \boldsymbol{0}|\boldsymbol{h}) = \sigma(W \boldsymbol{h} + \boldsymbol{a}),
\end{eqnarray*}
where $\sigma$ denotes the sigmoid function. This algorithm allows one to sample from the probability distribution defined by the amplitude RBM. The phase RBM does not define a probability distribution and hence requires no sampling.

We use only one CD step in each training epoch. Unique vectors are chosen among all samples. The log-likelihood function is estimated [according to Eq.~(3) in the main text based on the state \eqref{psiRBM} above] in each epoch using only these unique samples, weighted by their probabilities \eqref{prvRBM}. In the first training epoch, we initialize $\boldsymbol{s}$  with a set of $L$ random bit strings. On each subsequent epoch, CD is initialized with the CD output samples from the previous epoch  (a method known as persistent CD~\cite{tieleman2008training}).

\section*{Autoregressive neural network}\label{sec:model} 

\begin{figure}
\includegraphics[scale=0.5]{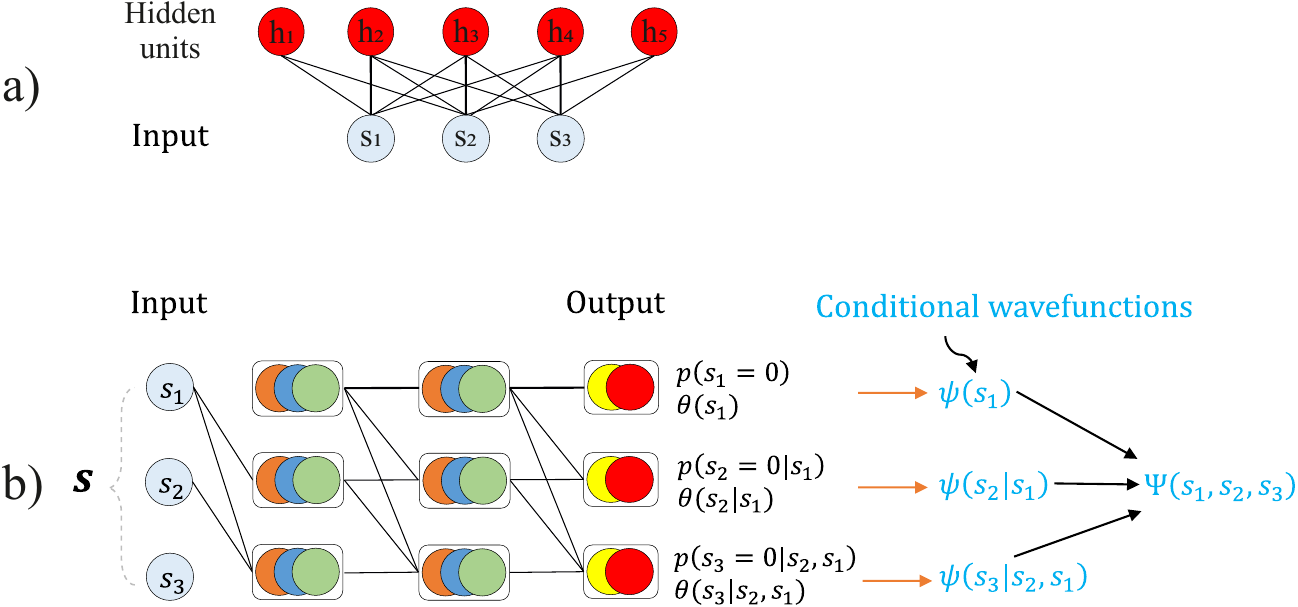}
\caption{Schematics of the neural networks used in this work for quantum variational optimization. a) RBM b) ARN. 
\label{fig:model}}
\label{fig:NNQS}
\end{figure}

The autoregressive model [Fig.~S\ref{fig:model}(b)] relies upon a simple observation that any joint probability distribution can be written as a product of one-dimensional nested conditional probabilities. That is, we can write $p(\boldsymbol{s})=\prod_{i=1}^N p\left(s_{i} | \boldsymbol{s}_{<i}\right)$, where $\boldsymbol{s}_{<i}=(s_1,\dots,s_{i-1})$. The output layer of a (classical) ARN consists of $N$ units, predicting $p\left(s_{i}=0 | \boldsymbol{s}_{<i}\right)$ for all $i$'s. To enable the computation of these units, the  connections of the $i$th output node are limited to those input units whose index is less than $i$. The hidden layers of the ARN are represented by clusters of multiple units corresponding to each qubit to achieve better expressivity of the network \cite{nade, papamakarios2017masked}.  The output layer utilizes sigmoid activation to ensure output values in the range of $\left(0,1\right)$. 

In the sampling regime, the network operates as follows [Fig.~S\ref{fig:NNQS}(b)]. One first computes $p(s_1=0)$ from the network parameters and samples $s_1$ accordingly. Subsequently, one computes $p(s_2=0|s_1)$ from the sampled $s_1$ and samples $s_2$. The process continues until all $s_i$ are sampled. In this way, the bits are sampled one at a time, each time the random choice being made out of 2 rather than $2^N$ options. This greatly simplifies the sampling process.

The above scheme can be readily adjusted to account for the quantum phases, as demonstrated by Sharir {\it et al.} \cite{Carleo2020}, where an ARN with convolutional features has been used to find the ground state of the transverse-field Ising model on a square lattice. In a quantum ARN, each output unit is augmented to include the conditional phase $\hat{\theta}(s_i)=\theta{\left(s_{i} | \boldsymbol{s}_{<i}\right)}$,  
so that 
\begin{equation}\label{wave_conditional}
\Psi(\boldsymbol{s}) 
= \prod_{i=1}^N \psi\left(s_{i} | \boldsymbol{s}_{<i}\right)
\end{equation}
with 
\begin{equation}\label{parameterization}
    \psi\left(s_{i} | \boldsymbol{s}_{<i}\right) =
    \sqrt{\hat{p}(s_i)} \exp [2\pi i \hat{\theta}(s_i)].
\end{equation}



\section{Selecting optimal model parameters}
\begin{figure}
\includegraphics[scale=0.5]{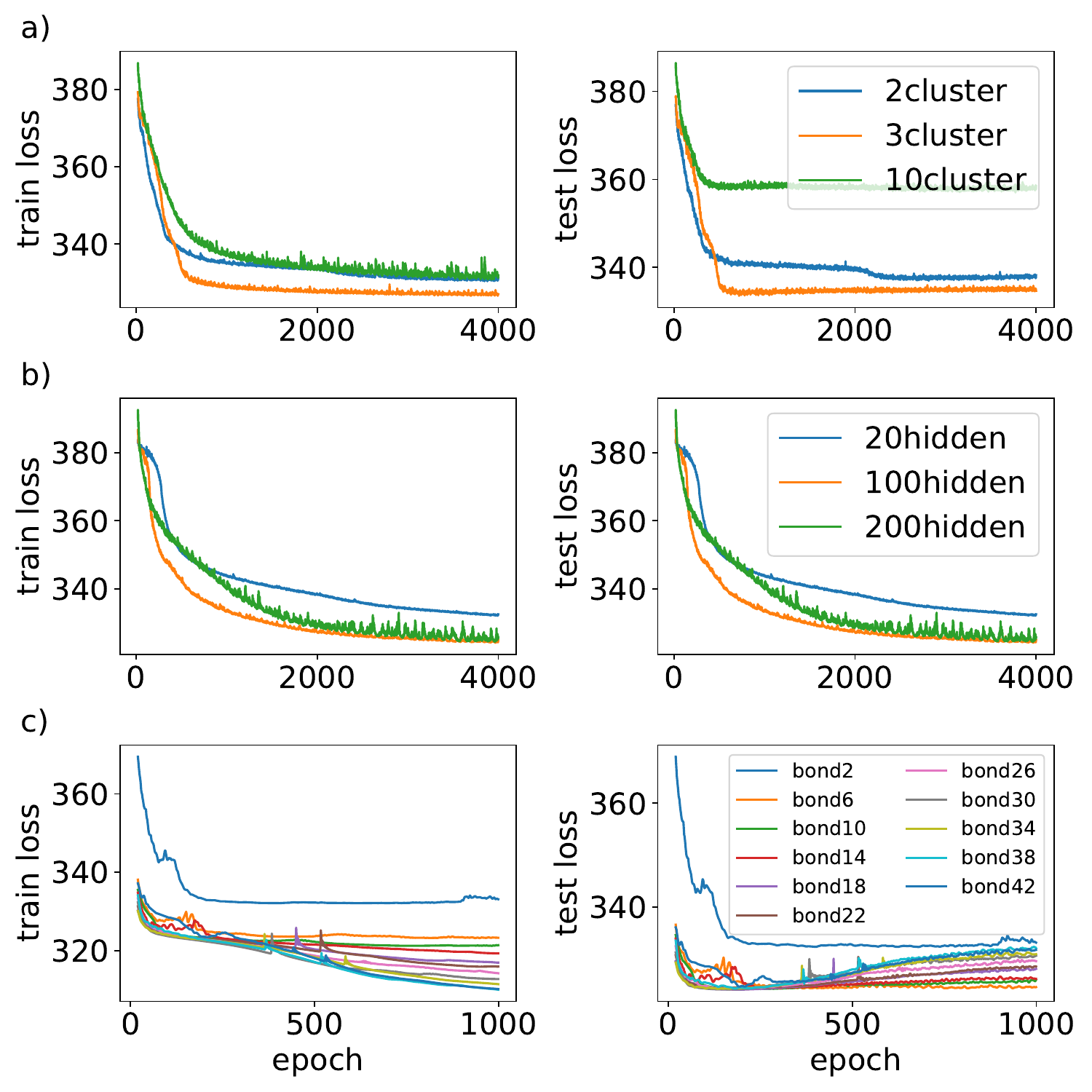}
\caption{Performance of different models with different number of parameters: a) ARN b) RBM c) MPS. Left: training; right: test.}
\label{fig:finding_optimal_model}
\end{figure}

To determine the optimal number of parameters for state reconstruction using the MPS, ARN, and RBM models, we conducted training using a range of model sizes. The MPS model was trained using bond dimensions ranging from 2 to 42 in increments of 4. The ARN model, containing two hidden layers, was trained with cluster sizes of 2, 3, and 10. The RBM model was trained using 20, 100, and 200 nodes in the hidden layer.


Figure~S\ref{fig:finding_optimal_model} presents the training and test loss for ARN, RBM, and MPS models with varying numbers of model parameters for the measurement data corresponding to state at $t=2.5$ ms. For the ARN model, the architecture with 3 units per cluster yielded optimal train and test loss, while for the RBM model, the architecture with 100 hidden nodes showed the best performance. The total number of model parameters for these two models is approximately 4000. For MPS, the performance, as evaluated by the minimum loss on the test set, saturates at the bond dimension of 10, which corresponds to approximately the same number of parameters.

\section{Effect of number of bases }

\begin{figure}[bt!]
 \includegraphics[width=0.45\textwidth]{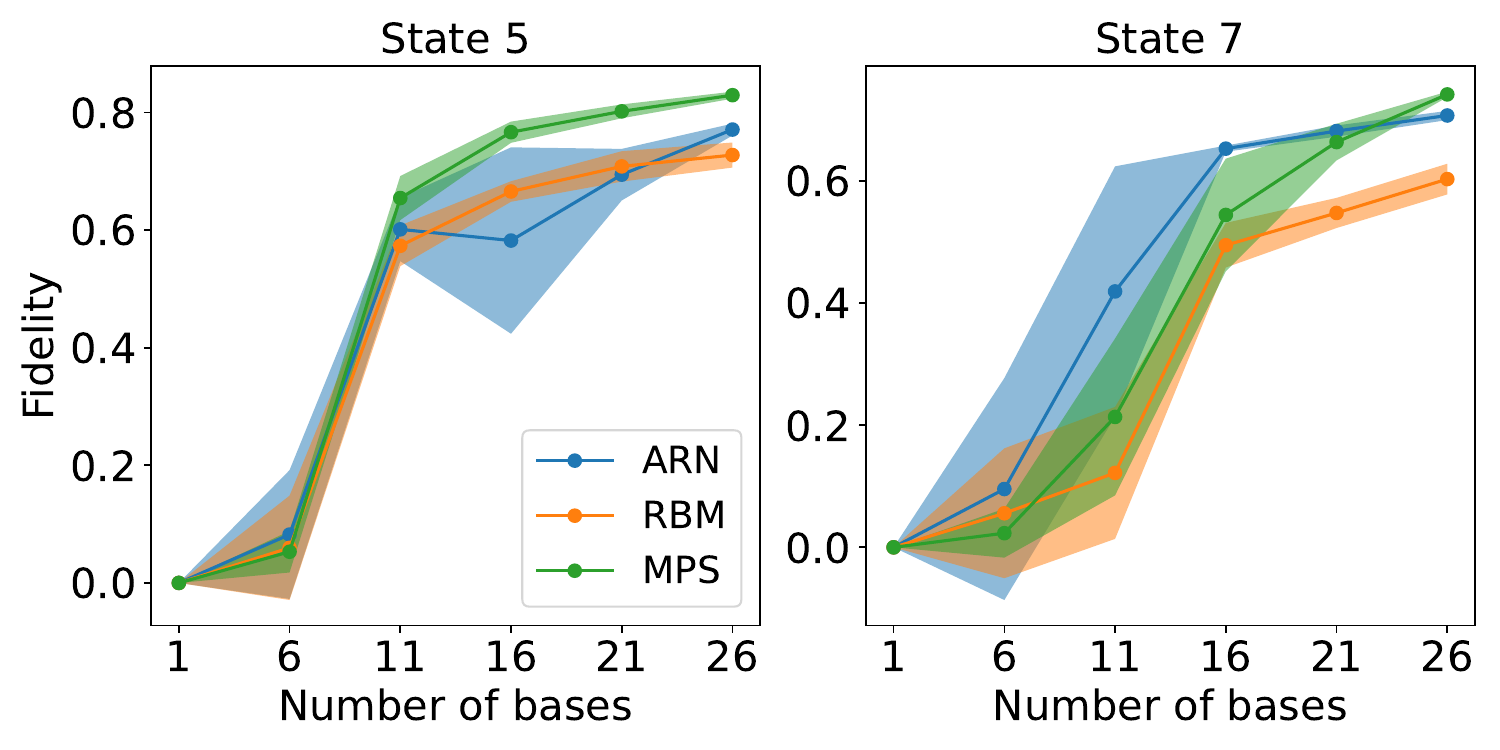}
 \caption{Effect of the number of measurement bases. The horizontal axis represents the number of bases used for the reconstruction taken randomly from the set of 27 bases. The vertical axis represents the fidelity between the full measurement set reconstruction and the reconstruction from the restricted set. Each point represents an average fidelity calculated using 5 reconstructions, and the shaded region represents one standard deviation. 
 }
 \label{fig:remove_bases}
\end{figure}

To further investigate the optimality of the setting chosen for the state tomography, we performed reconstructions using a reduced of bases chosen randomly from the 27 measurement bases and calculated the reconstructed fidelity with the theoretical wavefunction. The batch size $B$ is taken to be 150,000.  
The reconstructed fidelity increases with the number of bases we used in the reconstruction, with the increase slowing down when this number reached about 20. 
Our observations are illustrated in Fig.~S\ref{fig:remove_bases}.

\section{Effect of small additive in log-likelihood}
The expression for the loss (negative log-likelihood) used for the reconstruction contains a small additive $\varepsilon$:
\begin{equation}\label{LL}
	\Xi = -\sum_{p=1}^K\sum_{m=1}^{M_p}\ln\left(\abs{\langle\psi_m^p|\Psi_W\rangle}^2+\varepsilon \right).
\end{equation}
This is necessary to account for experimental events that have zero probability given the state $\ket{\Psi_W}$ --- for example, when this state predicts a qubit in the state $\ket 0$, but some measurements of that qubit in the computational basis produce $\ket 1$. Without $\varepsilon$, the log-likelihood of such a dataset would be negative infinity. If iterations start from such a state, the loss gradient cannot be computed and iterations cannot proceed. Setting a small positive  $\varepsilon$ helps eliminating this problem.

\begin{figure}[bt!]
 \includegraphics[width=0.75\textwidth]{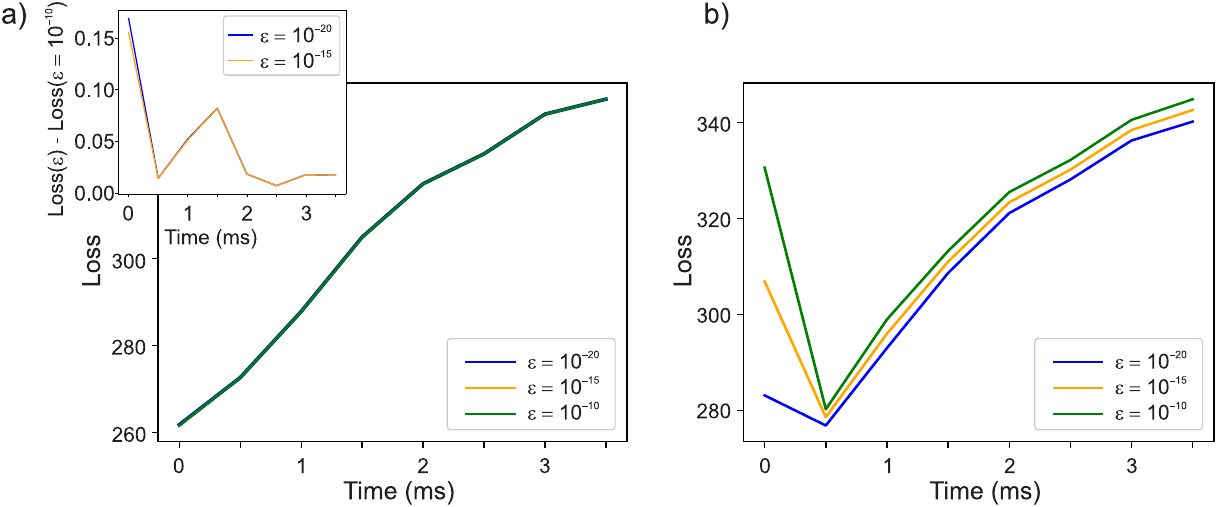}
 \caption{Loss on the test data for
the (a) experimentally reconstructed  via MPS and (b) theoretically predicted
states with different values of $\varepsilon$ in Eq.~\eqref{LL}. The three curves in (a) are indistinguishable; their minute difference is shown in the inset. The loss value converges with $\varepsilon\to 0$ for the experimentally reconstructed states, but not for theoretical ones.}
 \label{fig:epsilon} 
\end{figure}

In Fig.~S\ref{fig:epsilon}(a) we show the loss for the experimentally reconstructed state for various small values of  $\varepsilon$. We observe no significant dependence for $\varepsilon\lesssim10^{-10}$.

Interestingly, this is not the case if the loss \eqref{LL} is calculated for the theoretical state [Eq.~(2) in the main text], as can be seen in Fig.~S\ref{fig:epsilon}(b). This is likely because the theoretical states, for all times $t$, contain predictions that are incompatible with the observed measurement results. An example is the calculation of the loss with respect to the theoretical state (a  N{\`e}el-ordered product initial state $|\Psi(0)\rangle=|1, 0,1, \ldots\rangle$) at $t=0$. About 15\% of state preparation events suffer from single bit flip errors and an additional 2\% from double or higher-order bit flip errors. As a result, there is a nonzero probability of a theoretically impossible detection of e.g.~the first qubit in state $\ket 0$.



\section{Statistical uncertainty in the reconstructions}

\begin{figure}[bt!]
 \includegraphics[width=1\textwidth]{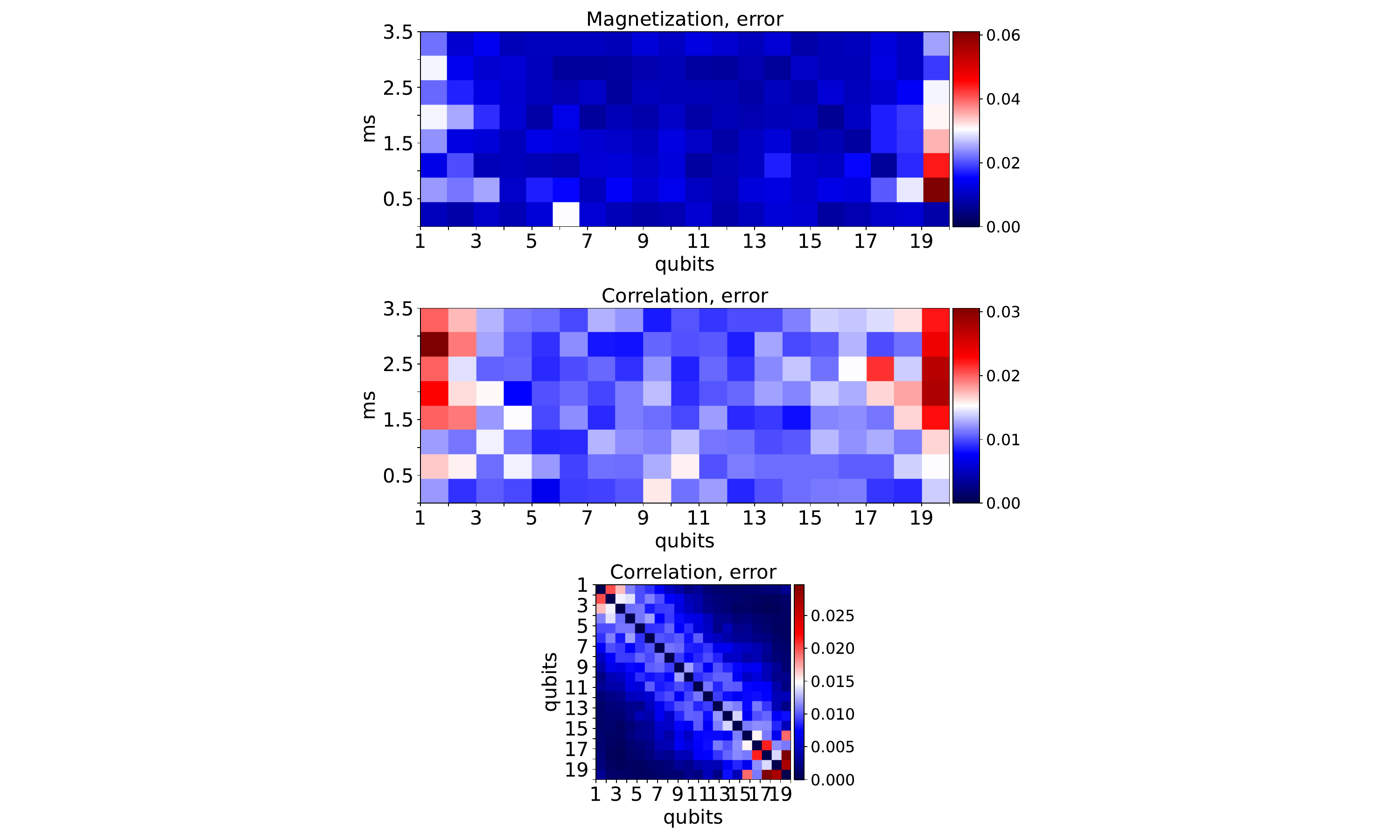}
 \caption{Standard deviation of the observables plotted in Fig.~1 of the main text, computed from \teal{twenty} bootstrapped reconstructions.}
 \label{fig: statistical_uncertainty} 
\end{figure}

By using the bootstrapping technique, one can assess the statistical uncertainty associated with the reconstruction. This approach involves generating multiple simulated sets of measurement data based on the reconstructed state. Subsequently, one can use any specific tomography scheme to perform reconstructions on each of these datasets. In Figure S\ref{fig: statistical_uncertainty}, we use this method to compute  the uncertainty. For each of the eight values of $t$, we produce \teal{twenty} simulated datasets from the experimental MPS-reconstructed states. We then calculate the standard deviations of the magnetization $\langle Y_i\rangle$ and correlation $\langle Y_iY_j\rangle$ observables of the secondary states reconstructed from these datasets. One can see that the deviations from the theoretical states observed in Fig.~1 of the main text are statistically significant. For example, the rms statistical error in the magnetization is on a scale of \teal{0.03 with a 30\% uncertainty}\footnote{\teal{The statistical error margin of estimating the width $\sigma$ of a Gaussian distribution from $N$ samples is  $\sigma\sqrt{2/N}$ \cite{bar2001estimation}.}}, whereas the difference between the reconstructed state and the theory in Fig.~1 is on a scale of 0.1--0.2 for some points.

\bibliography{references.bib}